# Moiré Hyperbolic Metasurfaces

Guangwei Hu[1,2], Alex Krasnok[2,3], Yarden Mazor[3], Cheng-Wei Qiu[1, †], Andrea Alù[2,3,4,5, †]

1. Department of Electrical and Computer Engineering, National University of Singapore, Singapore 117583, Singapore
2. Photonics Initiative, Advanced Science Research Center, City University of New York, 85 St. Nicholas Terrace, New York, NY 10031, USA
3. Department of Electrical and Computer Engineering, The University of Texas at Austin, Austin, Texas 78712, USA
4. Physics Program, Graduate Center, City University of New York, New York, NY 10016, USA
5. Department of Electrical Engineering, City College of New York, New York, NY 10031, USA

† Corresponding authors: aalu@gc.cuny.edu, chengwei.qiu@nus.edu.sg

**Recent advances in twistronics of low-dimensional materials, such as bilayer graphene and transition-metal dichalcogenides, have enabled a plethora of unusual phenomena associated with moiré physics. However, several of these effects require demanding manipulation of superlattices at the atomic scale, such as the careful control of rotation angle between two closely spaced atomic lattices. Here, we study moiré hyperbolic plasmons in pairs of hyperbolic metasurfaces (HMTSs), unveiling analogous phenomena at the mesoscopic scale. HMTSs are known to support confined surface waves collimated towards specific directions determined by the metasurface dispersion. By rotating two evanescently coupled HMTSs with respect to one another, we unveil rich dispersion engineering, topological transitions at *magic angles*, broadband field canalization, and plasmon spin-Hall phenomena. These findings open remarkable opportunities to advance metasurface optics, enriching it with moiré physics and twistronic concepts.**

**Kyewords:** *hyperbolic metasurface; moiré physics; twistronics; topological transition; graphene plasmons*

Recently, moiré superlattices have been successfully demonstrated in bilayer or hetero-bilayer two-dimensional (2D) materials, in which one layer is rotated with respect to the other, unveiling phenomena such as flatband superconductivity at a *magic* rotation angle[1] and atomic photonic crystals[2,3] in twisted bilayer graphene, topological excitons in stacked transition-metal dichalcogenides[4,5,6,7], and more[8,9,10]. The rich physics behind these discoveries is governed by the mediation of carrier potentials through van der Waals interactions and interlayer hopping energy between tightly coupled atomic lattices. Such interlayer coupling, as well as the hopping energy, are strongly dependent on the atomic lattice arrangement and thus, once rotated with respect to another, the moiré periodic pattern and its associated electronic bandstructure support significantly different electronic properties from those of each individual monolayer. The strong dependence on the rotation angle has opened the emerging research direction of "twistronics"[10]. Since the lattice constant of each monolayer is in the atomic scale range, interesting effects are achieved for very small rotation angles, producing moiré periodicities at larger scales. This requirement hinders practical opportunities in the context of moiré physics. Inspired by the twistronic concepts, we explore moiré hyperbolic metasurfaces (HMTSs) with significantly larger lattice granularity, which implies a broader range of relevant angles for which interesting phenomena can arise. The larger periods allow engaging not only the electronic bandstructure, but also photonic dispersion engineering.

Optical metasurfaces have been widely studied in recent years as the planarized version of metamaterials. Multiple anisotropic metasurfaces, stacked to form compact multifunctional metadevices, are typically designed with the goal of minimizing the coupling between adjacent layers[11,12,13,14,15,16], which drastically helps the design. However, near-field interactions of closely spaced metasurfaces open new opportunities, as we show in the following. In particular, we explore moiré metasurfaces obtained by stacked uniaxial metasurfaces supporting highly confined hyperbolic plasmons. The hyperbolic dispersion of HMTSs forces plasmon polaritons to propagate in collimated directions. In turn, this feature provides opportunities to control the evanescent coupling of two closely spaced HMTSs through the mediation of the rotation angle between them. Our explorations reveal the possibilities of dispersion engineering of the plasmon propagation in these structures, hyperbolic-to-elliptical topological transitions[17] at a *magic rotation angle*, flat band features with low-loss field canalization and tailored chirality. These findings extend "twistronics", topological bandstructure engineering and moiré physics to the domain of metasurfaces, making it relevant for the broad class of photonic devices.

**Moiré metasurfaces**

HMTSs are planar patterned surfaces that support in-plane propagation with hyperbolic dispersion. They combine the unusual features of hyperbolic metamaterials, such as broadband enhancement of local density of states, canalization and sub-diffraction imaging and negative refraction, with the advantage of supporting these exotic phenomena over a surface, not in the bulk, hence with more resilience to loss and easier accessibility. HMTSs have been demonstrated in several material platforms, from silver gratings[18], nanostructured van der Waals materials[19], graphene nanoribbons[20], black phosphorus,[21] and other 2D materials[22]. In this work, we consider densely packed graphene nanoribbons as the platform of choice, but other metasurface geometries may be equally viable, as a function of the wavelength range of interest. A sketch is shown in the inset of Fig. 1a, where each graphene strip has width $W$ and an air gap $G$ separates neighboring strips. We assume deeply subwavelength periodicity $P = W + G \approx 0.005\lambda_0$, where $\lambda_0$ is the free-space wavelength, so that the metasurface response can be homogenized with the effective surface conductivity tensor[23]

$$\bar{\bar{\sigma}} = \begin{bmatrix} \sigma_{\alpha\alpha} & 0 \\ 0 & \sigma_{\beta\beta} \end{bmatrix} \approx \begin{bmatrix} \dfrac{W}{P}\sigma_G & 0 \\ 0 & \dfrac{W\sigma_G\sigma_C}{P\sigma_c + W\sigma_G} \end{bmatrix}. \tag{1}$$

Here, $\sigma_C = -\dfrac{2i\omega\varepsilon_0 P}{\pi}\ln\left[\dfrac{1}{\sin(\pi G/2P)}\right]$ is the effective strip conductivity taking into account near-field coupling and nonlocality, $\sigma_G$ is the pristine graphene conductivity given by the Kubo formula[24], ω is the radial frequency, and $\varepsilon_0$ is the free-space permittivity. Here, $\alpha$ and $\beta$ are defined perpendicular and parallel to periodical directions, respectively. In the hyperbolic regime, the metasurface acts as a metal for one transverse field polarization (Im[$\sigma_{\alpha\alpha}$]>0) and as a dielectric for the other one (Im[$\sigma_{\beta\beta}$]<0). The field distributions ($E_z$) and dispersion bands, obtained as the Fourier spectrum of the field excited by a broadband $z$-oriented dipole simulated using a full-wave commercial software, are shown in Figs. 1b-c, which indeed demonstrates collimated field profiles. We can then form a moiré HMTSs by coupling two identical HMTSs, as sketched in Fig. 1a, separated by a dielectric gap of width $d$. The rotation angles $\theta_1$ and $\theta_2$ between the principal direction of each metasurface **α** and the **x** direction determine the nature of the moiré metasurface. For simplicity, we assume the background and spacer materials to be free space, but asymmetric backgrounds can be easily captured generalizing the formulation in this work.

The moiré metasurface dispersion can be evaluated by finding the poles of the dyadic Green's function of the system[25]. We need to consider the complete set of both transverse-electric (TE) and transverse-magnetic (TM) modes, since they couple together due to the anisotropic nature of HMTSs and the rotation between the two surfaces. The dispersion equation is

$$\begin{vmatrix} A & B \\ C & D \end{vmatrix} = 0 \tag{2.a}$$

$$A = \begin{bmatrix} 2k_z/k_0 + \eta_0\sigma_{hh,2} & \eta_0\sigma_{h\rho,2} \\ \eta_0\sigma_{\rho h,2} & 2k_0/k_z + \eta_0\sigma_{\rho\rho,2} \end{bmatrix} \tag{2.b}$$

$$\mathbf{B} = \begin{bmatrix} \eta_0 \sigma_{hh,1} & \eta_0 \sigma_{h\rho,1} \\ -\eta_0 \sigma_{\rho h,1} & -\eta_0 \sigma_{\rho\rho,1} \end{bmatrix} \qquad (2.c)$$

$$\mathbf{C} = \begin{bmatrix} \eta_0 \sigma_{hh,2} e^{2ik_z d} & \eta_0 \sigma_{h\rho,2}\, e^{2ik_z d} \\ \eta_0 \sigma_{\rho h,2} e^{2ik_z d} & \eta_0 \sigma_{\rho\rho,2} e^{2ik_z d} \end{bmatrix} \qquad (2.d)$$

$$\mathbf{D} = \begin{bmatrix} (2k_z/k_0 + \eta_0 \sigma_{hh,1}) & \eta_0 \sigma_{h\rho,1} \\ \eta_0 \sigma_{\rho h,1} & 2k_0/k_z + \eta_0 \sigma_{\rho\rho,1} \end{bmatrix}. \qquad (2.e)$$

Here, $k_z^2 = k_0^2 - k_\rho^2$ is the z-component of the wavevector in vacuum, associated with the in-plane wavevector $k_\rho \hat{\rho} = k_x \hat{x} + k_y \hat{y}$. $\sigma_{ij,\gamma}, i,j \in \{\rho, h\}, \gamma \in \{1,2\}$ are the components of the surface conductivity tensor of the first and second HMTSs projected along parallel ($\hat{\rho}$) and perpendicular ($\hat{h}$) directions of the in-plane wavevectors. The diagonal sub-matrices $\mathbf{A}$, $\mathbf{D}$ determine the dispersion of the first and second HMTS when isolated, consistent with previous results[20,21,22,23]. The off-diagonal elements $\mathbf{B}$, $\mathbf{C}$ model the coupling between them. $\mathbf{C}$ goes to zero as $d$ grows to infinity and the two HMTSs become uncoupled.

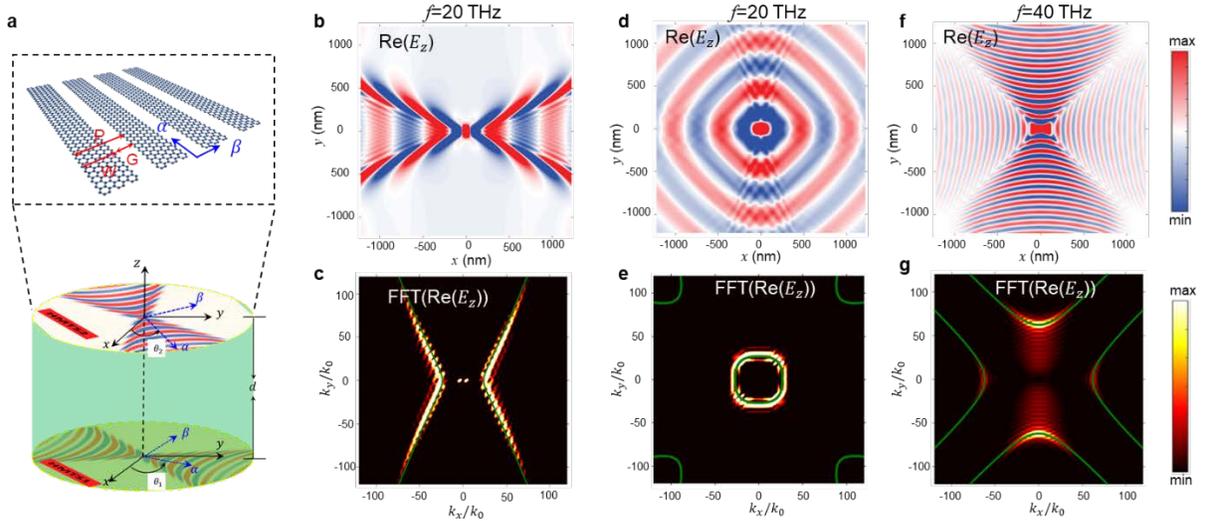

Figure.1 **Geometry, field distributions and analytical dispersion of moiré hyperbolic metasurfaces.** |(a) Geometry of moiré hyperbolic metastructures, composed of two coupled uniaxial metasurfaces with different rotation angles $\theta_1$ and $\theta_2$ with respect to $\boldsymbol{x}$. Their prime characteristic directions are denoted as $\alpha$ and $\beta$. Two HMTSs are separated by a dielectric spacer and thickness $d$. (b). $E_z$ distribution for an individual HMTSs in vacuum and (c) its dispersion at frequency $f$=20THz. (d) $E_z$ distribution for the moiré

HMTS and (e) its dispersion at frequency $f$=20THz. (f) $E_z$ distribution for the moiré HMTS and (g) its dispersion at $f$=40THz. The green curve in c, e and g are the numerically calculated dispersion. HMTSs in (b) and (c) are composed by graphene ribbons ($P$=50nm, $W$=30nm, Fermi level = 0.4eV and $\tau = 0.5$ps). The moiré HMTS in (d) and (e) is composed of the same graphene ribbons with $\theta_1 = 0$, $\theta_2 = 90°$ and $d = 10$nm.

**Topological Transition *Magic Angles***

Graphene metasurfaces enable a large degree of tunability via chemical doping of the hyperbolic dispersion[20]. Here, unless otherwise explicitly stated, we fix the Fermi level to 0.4 eV and choose the other geometrical parameters to be $P$=50nm, $W$=30nm, and $\theta_1$=0° throughout the work. To study the effect of stacking, we first study the simple scenario of twisted bilayer nanoribbons with $\theta_2 = 90°$ and $d$=10nm. Fig. 1d,f show the field distribution at 20 THz and 40 THz, with isofrequency contours given in Fig. 1e,g. An excellent agreement between numerical and analytical results is observed. Despite the hyperbolic nature of the two surfaces when isolated at both frequencies, the moiré stack supports elliptical contours at 20 THz and hyperbolic ones at 40 THz, hinting to exciting opportunities of dispersion engineering enabled by stacking.

To thoroughly explore these effects, Figs. 2a-b show the isofrequency dispersion curves for moiré HMTSs considering different rotation angles at 20 THz and 40 THz, respectively. To a very good extent, we can predict the overall response of the stack as the superposition of the dispersion of the individual metasurfaces suitably rotated one from the other. At the points in which the two hyperbolic bands intersect, we find peculiar anti-crossing features in the isofrequency contours. The anti-crossing is visible also in full-wave simulations, even though the high-$k$ branches are poorly excited by localized emitters, as seen in Fig. 1e from the weak hyperbolic branches in the corners. We observe that at 20THz and $\theta_2$=60°, as well as $\theta_2$=90° (consistent with Fig. 1e), the close interaction of the two surfaces forms a closed dispersion band at small $k$ values, which however opens and becomes hyperbolic for $\theta_2=30°$, experiencing a topological transition near 45°. The closed small-$k$ branch corresponds to a hybrid TE-TM mode, and it does not arise at 40 THz, regardless of the rotation angle. Such topological transition is the analogue of a *Lifshitz transition* in electronic bandstructures[26], playing a crucial role in the physics of Weyl and Dirac semimetals[27]. In optics, this phenomenon has been predicted in graphene hyperbolic metasurfaces and it can be unveiled by tuning the chemical potential,[20] or sweeping the frequency[17]. Our findings suggest the opportunity of inducing topological transitions for fixed frequency and bias by tuning the rotation angle of two HMTSs.

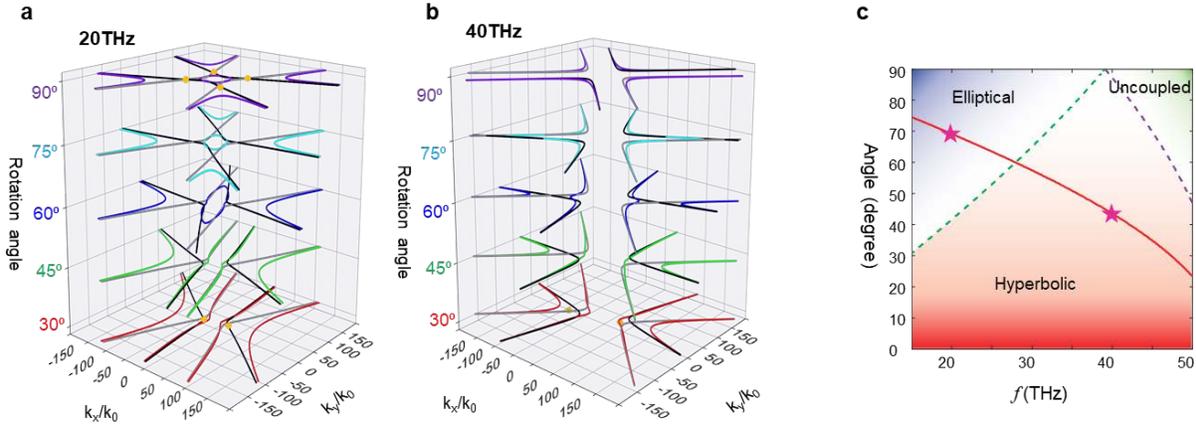

Figure 2 **Topological transition magic angles** | (a)-(b). Dispersion as a function of the twisted angle at 20THz, and 40 THz. The red, green, blue, cyan and purple solid lines denote the dispersion of moiré HMTSs with the rotation angle of 30˚, 45˚, 60˚, 75˚ and 90˚, respectively. The grey (black) solid lines at each plot denote the dispersion of the first (second) individual HMTSs, where the yellow dots mean the anti-crossing points of dispersion in two individual HMTSs. (c). The topological transition regions as a function of frequency and rotation angles. The red solid line represents the open-angle ($\psi$) of the first HMTS, with respect to the frequency. The green dashed line represents the second topological transition angle ($180° − 2\psi$) and the purple dashed lines represent the first topological transition angle ($2\psi$). The pink 5-point stars denote open angles at 20 THz and 40 THz, which are 69.1° and 43.6°, respectively. The blue region corresponds to the presence of four anti-crossing points (elliptical regime, see dispersions of $\Delta\theta=60°$, 75˚ and 90˚ at 20 THz), the red region indicates two anti-crossing points (hyperbolic regime, see dispersions of $\Delta\theta=30°$ and 45˚ at 20 THz as well as $\Delta\theta=30°$, 45˚ and 60˚ at 40 THz) and the green area indicates no anti-crossing (uncoupled regime, see dispersion of $\Delta\theta=90°$ at 40 THz).

The observed topological transition can be powerfully modeled and predicted considering the open angle of the induvial hyperbolic branches ψ and the rotation angle between them ($\Delta\theta = \theta_2 - \theta_1 \in [0, 90°]$). The open angle ψ (red solid curve in Fig.2c) of the individual HMTS is defined as $\tan\psi = \sqrt{-\frac{\text{Im}(\sigma_{\alpha\alpha})}{\text{Im}(\sigma_{\beta\beta})}}$.[20] When the surfaces are paired together with a given rotation angle, an anti-crossing arises at the intersection of the bands (see Fig. S2 for more details), which may be easily predicted using geometrical arguments. We do not expect anti-crossing if $2\psi < \Delta\theta < 180° - 2\psi$, and in this case the HMTSs are weakly coupled, with only slight modifications of the isolated dispersion (see $\theta_2$=90° at 40 THz in Fig. 2b). On the contrary, two anti-crossing points emerge in the first and third quadrants if $\theta < 2\psi$ and $\theta < 180° - 2\psi$ and, as a result, an open small-$k$ branch arises. Four anti-crossing points arise if $180° - 2\psi < \Delta\theta < 2\psi$, and the small-$k$ branch will be essentially closed (see $\theta_2$=60° and $\theta_2$=90° at 20 THz in Fig. 2b). The different regimes are sketched in Fig. 2c, and are consistent with the rigorously calculated bands in Fig. 2a-b. The special condition to achieve a topological transition is obtained when $\Delta\theta = 2\psi$, if the open angle of the individual HMTS dispersion satisfies $\psi < 45°$ (dashed purple line in Fig. 2c), and a second topological angle $\Delta\theta = 180° - 2\psi$ (dashed green line in Fig. 2c) exists if $\psi > 45°$. Near the second topological transition angle ($\Delta\theta = 180° - 2\psi$) we find that the small-$k$ branch turns into nearly parallel straight lines (see $\Delta\theta$=45º in Fig. 2a and more details in Fig. S3), indicating that the surface plasmons in this regime are canalized and propagate at a fixed direction. Our full-wave simulations also confirm this transition behavior in a realistic double-layer structure of graphene nanoribbons, Fig. S8 and Fig. S9.

**Broadband Field Canalization**

Since the discovery of hyperbolic metamaterials, flat band dispersion has been one of the particularly appealing properties of this class of materials, ideally suited for imaging purposes, as it allows to canalize the fields along the desired direction without diffraction, known as field canalization[28]. Hyperlenses for subwavelength imaging[23,29] and spontaneous emission rate enhancement[20,23] have been proposed in this regime based on various platforms, e.g., capacitively-loaded wire media[28,30], where the band flattens at the vicinity of band edges associated with zero group velocity, or HMTSs working at the condition of extreme anisotropy, such as conductivity-near-zero regimes[20,23]. However, these implementations only work at specific frequencies, and the damping near resonance significantly hinders their performance. In contrast, our results reveal the possibility of broadband flat dispersion in moiré metastructures out of resonance, owing to the controlled coupling of two HMTSs.

In order to prove this opportunity, we study the dispersion in the frequency range supporting four anti-crossing points, and we focus on $\Delta\theta = 90°$, Fig. 3a. As discussed above, the evanescent coupling induces band splitting and it results in a flattened small-$k$ branch due to the hybridization of TE and TM modes supported by the two HMTSs. At 15 THz, the isofrequency contour is circular, consistent with the TE mode of individual HMTSs. However, as the frequency increases to 20 THz and further to 28 THz, the dispersion flattens over a broad bandwidth. Such peculiar dispersion ensures that the Poynting vector and group velocity have fixed directions, independent of the wave number[31]. As a result, the field propagates towards the four directions perpendicular to the dispersion lines. Interestingly, these features arise far from the resonant frequencies of individual HMTSs usually associated with increased absorption. Thus, we can expect low-loss propagation of these canalized plasmons in our moiré hyperbolic metastructure.

To demonstrate this feature, we present full-wave simulations of the moiré structure in Figs. 3b-c. We observe an excellent agreement between analytical (green line) and numerical results (contour plot), Fig. 3b. The associated Poynting vector plotted in Fig. 3c indicates field canalization along *x* and *y*, perpendicular to the dispersion, as well as propagation distance five times longer than the plasmon wavelength ($2\pi/k_\rho$). The high-*k* branch, in contrast, is largely confined and poorly excited by a localized source in our full-wave simulations. Supplementary sections 4 and 8 show additional full-wave simulations of a realistic double-layer graphene nanoribbon metasurface, highlighting the broadband response of these phenomena. Overall, we find that the peculiar physics of the moiré metasurface enables unique dispersion engineering for broadband plasmon canalization.

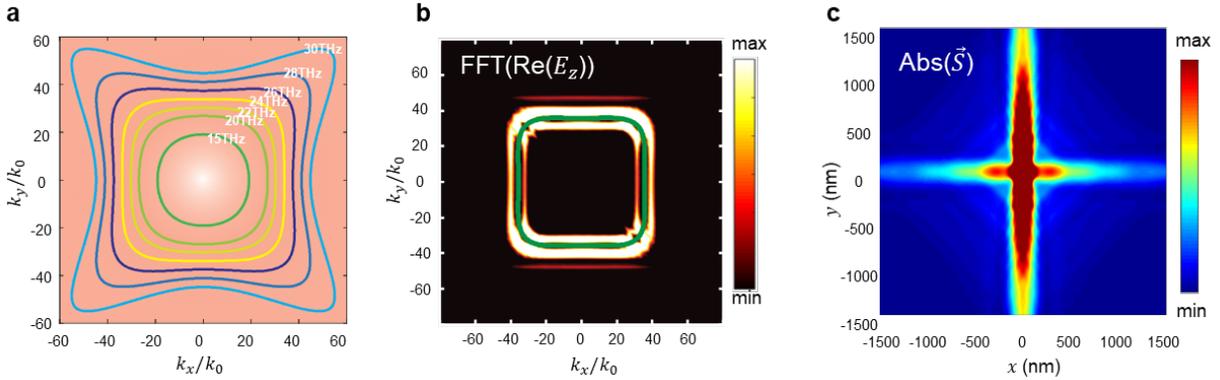

Figure 3 **Broadband field canalization** | (a). Dispersion of moiré hyperbolic metastructures at 15THz (dark green), 20THz (light green), 22THz (yellowish-green), 24THz (yellow), 26THz (purple), 28THz (dark blue) and 30THz (light blue). (b) Full-wave simulation of the dispersion of moiré HMTSs excited by a dipole polarized along *z* placed 70nm above the moiré stack. The twisted angle, in this case, is 90˚, *f*=25THz. The green solid line shows the analytical dispersion. Fermi level is 0.4eV. (c). Poynting vector ($\vec{S}$). Most of the energy goes towards *x* and *y*, showing energy canalization.

**Spin-Orbit Interactions**

Another remarkable opportunity of moiré metasurfaces is the emergence of spin-orbit interactions driven by the rotation angle. This response is analogous to the electron spin-orbit interaction that leads to valley degrees of freedom in twisted bilayer transition metal dichalcogenides, as a function of the symmetries governing the local atomic registry and the rotation angle[8]. Photonic spin-orbit interactions are of fundamental importance to realize multifunctional metadevices[18,32,33], working in near-field[34,35,36] and far-field[10,37]. Of particular interest are near-field interactions of a circularly polarized dipole close to an interface, enabling unidirectional launching of plasmons at metal-dielectric interfaces[38] and the plasmonic spin-Hall effect in HMTSs[18].

In our geometry, we can analyze the chirality of the field distribution analyzing the eigenmodes of the system, which consist of hybridized TE ($a_2$) and TM ($b_2$) modes:

$$\vec{E} \propto \tilde{p}_{yz}\hat{x} + i\hat{z}; \quad \tilde{p}_{yz} = \frac{k_0}{k_\rho C}\frac{1}{k_\rho}\left(k_x + C\frac{q}{k_0}k_y\right) \tag{3}$$

with

$$C = C_{\text{HMTS}} = \left(2 + \frac{k_0}{q}\frac{\eta_0}{k_\rho^2}\left(k_x^2 X_{\beta\beta} + k_y^2 X_{\alpha\alpha}\right)\right)\frac{k_\rho^2}{\eta_0 k_x k_y (X_{\alpha\alpha} - X_{\beta\beta})} \tag{4}$$

for individual HMTSs, and

$$C = C_M = -\frac{4\frac{q}{k_0} + 2\eta_0(X_{\alpha\alpha} + X_{\beta\beta}) + \frac{\eta_0^2}{qk_0}\frac{1}{k_\rho^2}\left\{k_0^2 k_\rho^2 X_{\alpha\alpha}X_{\beta\beta} + k_x^2 k_y^2 (X_{\alpha\alpha} - X_{\beta\beta})^2\right\}(1 - e^{-2qd})}{\eta_0^2 \frac{k_x k_y}{k_\rho^2}\frac{1}{k_0^2}\left\{k_y^2 X_{\alpha\alpha}^2 - k_x^2 X_{\beta\beta}^2 + (k_x^2 - k_y^2)X_{\alpha\alpha}X_{\beta\beta}\right\}(1 - e^{-2qd})} \tag{5}$$

for moiré hyperbolic metastructures with 90° rotation angles. Here, $k_z = iq = \sqrt{k_0^2 - k_\rho^2}$, $X_{\alpha\alpha} = \text{Im}(\sigma_{\alpha\alpha})$ and $X_{\beta\beta} = \text{Im}(\sigma_{\beta\beta})$. The coefficient $\tilde{p}_{yz}$, i.e., the complex ratio of the *y*- and *z*-eigenmode components measures its chirality. In the lossless case, $\tilde{p}_{yz}$ is purely real since $k_x$ and $k_y$ are real. As seen in Eq. 3, the eigenmodes indeed carry different out-of-plane chirality, which can couple selectively to circularly polarized emitters.

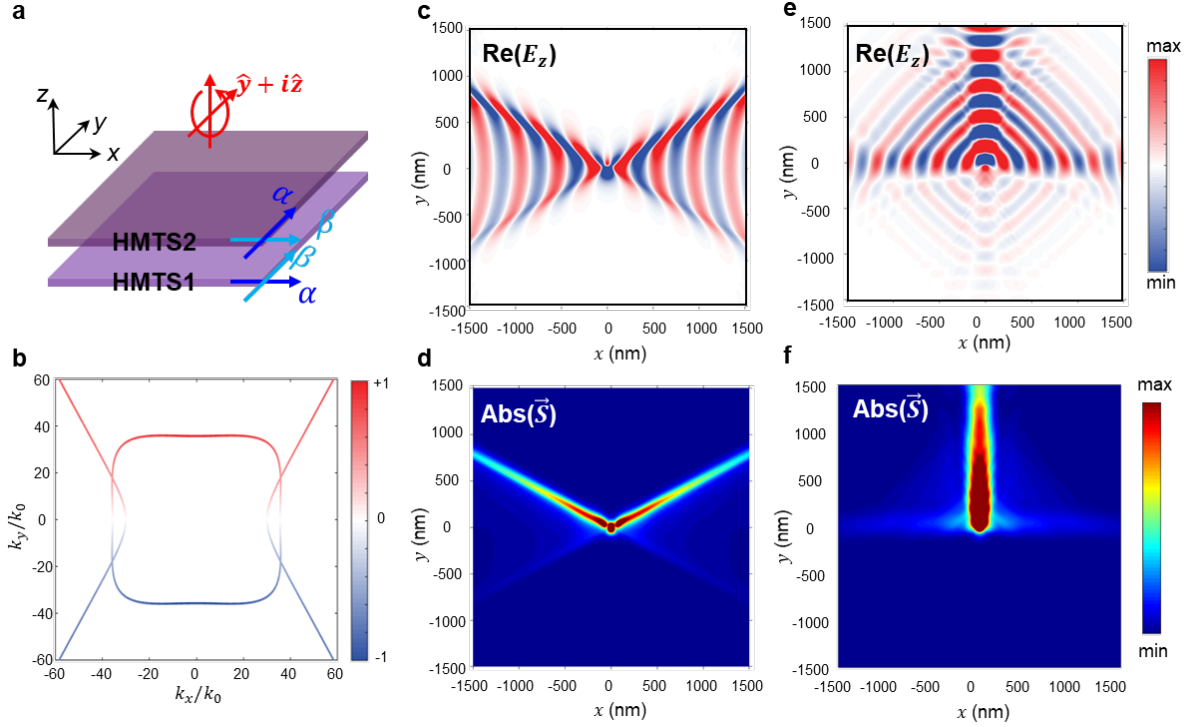

Figure 4 **Plasmon spin-Hall effect** | (a). Schematic illustration of moiré hyperbolic metasurfaces excited by an $\hat{y} + i\hat{z}$ dipole. In this scheme, the twisted angle is 90°. (b) Plasmon spin-Hall effect in moiré HMTSs based on the eigenmode analysis. Dispersion at 25THz of both individual HMTS and moiré metastructure. The color denotes the value of $\tilde{p}_{yz}$, which determines the circular polarization degree of the eigenmode, describing how the mode couples to circularly polarized dipoles $\tilde{p}_{yz}\hat{x} + i\hat{z}$. (c). Field distribution and (d) Poynting vector of modes of the individual HMTSs excited by the dipole $\hat{y} + i\hat{z}$ at 25 THz. (e). Field distribution and (f) Poynting vector for the moiré metastructures excited by the dipole $\hat{y} + i\hat{z}$ at 25 THz.

As an example, we use a circular-polarized dipole to excite the individual HMTSs and the moiré metastructure at 25 THz, Fig. 4a. The dispersion of individual HMTSs and the moiré metastructure are shown in Fig. 4b, with colors indicating the degree of chirality encoded in $\tilde{p}_{yz}$, as the eigenmode evolves along the contour lines. The modes at $k_y > 0$ mostly couple to $\hat{y} + i\hat{z}$, while the modes at $k_y < 0$ majorly couple to $-\hat{y} + i\hat{z}$, both for the individual HMTSs and for the moiré stack. This result ensures that we can endow the dispersion engineering schemes described in the previous sections, including the flat band properties, with strong spin-orbit coupling stemming from the HMTSs. To verify our analysis, we performed full-wave simulations for an $\hat{y} + i\hat{z}$ dipole excitation placed 70 nm above the moiré stack. The field distribution for individual HMTS and of the moiré metastructure are shown in Figs. 4c-f. For the individual HMTS, the plasmons selectively propagate towards the positive *y*-direction, with two energy flux channels following the two hyperbolic branches. In the moiré metastructure we observe similar selectivity in polarization, but a drastically modified dispersion diagram. In Supplementary Section 7 we discuss also the peculiar energy flow for chiral sources polarized in the other plane, $\pm\hat{x} + i\hat{z}$.

**Discussion and Conclusions**

In this work, we have examined the opportunities stemming from the rotation of two tightly coupled HMTSs in terms of dispersion engineering and topological transitions, enabling the emergence of exciting moiré physics for surface plasmon propagation. At specific topological transition angles, which may be predicted with simple geometric considerations studying the hyperbolic properties of the individual metasurfaces, two nearly parallel dispersion curves can be induced, enabling broadband field canalization with low damping. By tuning the chemical potential of graphene, we may further control the band dispersion of a given moiré stack with powerful tunability opportunities, as further discussed in supplementary section 5. We have also outlined the potential for strong spin-orbit interactions and plasmonic spin-Hall effect in these stacks, again controlled by the rotation angle. Although we have focused on pairs of identical HMTSs, even more exotic features may emerge considering different HMTSs. Supplementary section 6 studies the tunability of the response as the coupling distance is changed. The features obtained in this study may be experimentally implemented stacking and twisting HMTSs, such as graphene nanoribbons or subwavelength silver gratings at visible frequencies[18], double-layer nanostructured van der Waals materials[19], black phosphorus[21], or even natural hyperbolic materials. The proposed moiré hyperbolic metasurfaces based on the rotation-angle-mediated interlayer coupling of 2D HMTSs enable a new degree of spatial dispersion engineering, representing a promising platform to study moiré physics and twistronics in photonics, with opportunities for topologically tunable high-performance metadevices with desired functionalities.

**Corresponding Author:**

Cheng-Wei QIU: chengwei.qiu@nus.edu.sg

Andrea Alù: aalu@gc.cuny.edu

**Author contributions**

G. H., A. K., and A. A. conceived the idea. G. H., A. K., and Y. M. developed the theory of moiré metastructures. G.H. and A. K. performed the numerical calculations and full-wave simulations. All authors have analyzed and discussed the results. G. H. wrote the paper with the input of all authors. C.W.Q. and A. A. supervised the project.

**Acknowledgments**

This work was partially supported by the Air Force Office of Scientific Research and the National Science Foundation. C.-W.Q. acknowledges the financial support from the National Research Foundation, Prime Minister's Office, Singapore under its Competitive Research Program (CRP award NRF-CRP15-2015-03).

**Supporting Information**

Supporting information will be available online.

**Competing interests:** The authors declare no competing interests.

# Reference

1. Cao, Y. et al. Unconventional superconductivity in magic-angle graphene superlattices. *Nature* 556, 43–50 (2018).

2. Jiang, L., et al. Soliton-dependent plasmon reflection at bilayer graphene domain walls. *Nature materials* **15**, 840-844 (2016).

3. Sunku, S. S., et al. Photonic crystals for nano-light in moiré graphene superlattices. *Science* 362, 1153-1159 (2018)

4. Tran, K. et al. Evidence for moiré excitons in van der Waals heterostructures. *Nature* **567**, 71–75 (2019).

5. Seyler, K. L. et al. Signatures of moiré-trapped valley excitons in $MoSe_2/WSe_2$ heterobilayers. *Nature* **567**, 66–70 (2019).

6. Jin, C. et al. Observation of moiré excitons in $WSe_2/WS_2$ heterostructure superlattices. *Nature* **567**, 76–80 (2019).

7. Alexeev, E. M. et al. Resonantly hybridized excitons in moiré superlattices in van der Waals heterostructures. *Nature* **567**, 81–86 (2019).

8. Yu, H., Liu, G.-B., Tang, J., Xu, X. & Yao, W. Moiré excitons: From programmable quantum emitter arrays to spin-orbit–coupled artificial lattices. *Science Advances* **3**, e1701696 (2017).

9. Wu, F., Lovorn, T. & MacDonald, A. H. Topological exciton bands in moiré heterojunctions. *Phys. Rev. Lett.* **118**, 147401 (2017).

10. Carr, S. et al. Twistronics: Manipulating the electronic properties of two-dimensional layered structures through their twist angle. *Physical Review B* **95**, 075420 (2017).

11. Zhao, Y., Belkin, M. A. & Alù, A. Twisted optical metamaterials for planarized ultrathin broadband circular polarizers. *Nature Communications* **3**, 870 (2012).

12. Xu, H.X. et al. Chirality-Assisted High-Efficiency Metasurfaces with Independent Control of Phase, Amplitude, and Polarization. *Advanced Optical Materials* **7**, 1801479 (2019).

13. Xu, H.X. et al. Interference-assisted kaleidoscopic meta-plexer for arbitrary spin-wavefront manipulation. *Light: Science & Applications* **8**, 3 (2019).

14. Zhao, Y., Shi, J., Sun, L., Li, X. & Alù, A. Alignment-Free Three-Dimensional Optical Metamaterials. *Advanced Materials* **26**, 1439-1445, (2014).

15. Zhao, Y. et al. Chirality Detection of Single-Enantiomer Drugs Using Twisted Optical Metamaterials. *Nature Communications*, 14180, (2017).

16. Askarpour, A. N., Zhao, Y. & Alù, A. Wave Propagation in Twisted Metamaterials. *Physical Review B* **90**, 054305 (2014).

17. Krishnamoorthy, H. N. S., et al. Topological Transitions in Metamaterials. *Science* **336**, 205-209 (2012).

18. High, A. A. et al. Visible-frequency hyperbolic metasurface. *Nature* **522**, 192-196 (2015).

19. Li, P. et al. Infrared hyperbolic metasurface based on nanostructured van der Waals materials. *Science* **359**, 892-896 (2018).

20. Gomez-Diaz, J. S., Tymchenko, M. & Alù, A. Hyperbolic plasmons and topological transitions over uniaxial metasurfaces. *Physical Review Letters* **114**, 233901 (2015).

21. Correas-Serrano, D., Gomez-Diaz, J.S., Melcon, A.A. & Alù, A. Black phosphorus plasmonics: anisotropic elliptical propagation and nonlocality-induced canalization. *Journal of Optics* **18**, 104006 (2016).

22. Nemilentsau, A., Low, T. & Hanson, G. Anisotropic 2D materials for tunable hyperbolic plasmonics. *Physical Review Letters* **116**, 066804 (2016).

23. Gomez-Diaz, J. S. & Alù, A. Flatland optics with hyperbolic metasurfaces. *ACS Photonics* **3**, 2211-2224 (2016).

24. Chen, P.Y. & Alù, A. Atomically thin surface cloak using graphene monolayers. *ACS Nano* **5**, 5855-5863 (2011).

25. Gomez-Diaz, J.S., Tymchenko, M. & Alù, A. Hyperbolic metasurfaces: surface plasmons, light-matter interactions, and physical implementation using graphene strips. *Optical Materials Express* **5**, 2313-2329 (2015).

26. I. M. Lifshitz, Anomalies of electron characteristics of a metal in the high-pressure region. *Sov. Phys. JETP* 11, 1130 (1960).

27. Armitage, N.P., Mele E.J., Vishwanath A., Weyl and Dirac semimetals in three-dimensional solids. *Rev Mod Phys* **90**, 015001 (2018).

28. Belov, P.A., Simovski, C.R. & Ikonen, P. Canalization of subwavelength images by electromagnetic crystals. *Physical Review B* **71**, 193105 (2005).

29. Li, Z.-Y., & Lin, L.-L. Evaluation of lensing in photonic crystal slabs exhibiting negative refraction. *Physical Review B* **68**, 245110 (2003).

30. Shvets, G., Trendafilov, S., Pendry, J.B. & Sarychev, A. Guiding, focusing, and sensing on the subwavelength scale using metallic wire arrays. *Physical Review Letters* **99**, 053903 (2007).

31. Ferrari, L., Wu, C., Lepage, D., Zhang, X. & Liu, Z. Hyperbolic metamaterials and their applications. *Progress in Quantum Electronics* **40**, 1-40 (2015).

32. Bliokh, K. Y., Rodríguez-Fortuño, F. J., Nori, F. & Zayats, A. V. Spin–orbit interactions of light. *Nature Photonics* **9**, 796–808 (2015).

33. Zhou, J. et al. Broadband photonic spin Hall meta-Lens. *ACS Nano* **12**, 82–88 (2018).

34. Lin, J. et al. Polarization-controlled tunable directional coupling of surface plasmon polaritons. *Science* **340**, 331–334 (2013).

35. Tsai, W.Y. et al. Twisted Surface Plasmons with Spin-Controlled Gold Surfaces. *Advanced Optical Materials*, 1801060 (2019).

36. Sun, L. et al. Separation of Valley Excitons in a $MoS_2$ Monolayer Using a Subwavelength Asymmewtric Groove Array *Nature Photonics* **13**, 180-184 (2019).

37. Hu, G. et al. Coherent steering of nonlinear chiral valley photons with a synthetic Au–$WS_2$ metasurface. *Nature Photonics* **13**, 467-472 (2019).

38. Rodríguez-Fortuño, F.J. et al. Near-field interference for the unidirectional excitation of electromagnetic guided modes. *Science*, **340**, 328-330 (2013).